\documentclass[12pt]{iopart}

\usepackage{color}
\usepackage{graphicx}
\usepackage{iopams}
\usepackage{setstack}
\begin{document}

\title[HS under shear flow]{Hydrodynamic interactions of colloidal spheres under shear flow}

\author{M Radu$^{1,2}$ and T Schilling$^1$}

\address{$^1$ Theory of Soft Condensed Matter, Universit{\'e} du Luxembourg, L-1511 Luxembourg, Luxembourg\\
		$^2$ Institut f{\"u}r Physik, Johannes Gutenberg Universit{\"a}t, D 55099, Mainz, Germany}
\ead{marc.radu@uni.lu}
\begin{abstract}
Particles that are immersed in a fluid exchange momentum via the fluid, hence 
their Brownian motion is correlated.   
By means of multiparticle-collision dynamics 
simulations we study the interactions between two colloidal beads in a 
sheared fluid suspension. Recently, this topic has been addressed in 
experiments on colloidal particles 
trapped by optical tweezers in a microfluidic device 
[PRL {\bf 103}, 230602 (2009)] and theoretically by means of a Langevin model [Eur.~Phys.~J E {\bf 33}, 313 (2010)]. 
Although we neglect the rotational degrees of freedom of the colloids,  
and employ a very simple coupling between the 
colloids and the flow field, we can reproduce the experimental data and
partly explain why it differs from theory.
\end{abstract}

\submitto{\JPCM}
\maketitle

\section{Introduction\label{sec:Intro}}
A mesoscopic particle that is suspended in a liquid moves stochastically 
due to collisions with the particles of the 
liquid -- it performs Brownian motion. When two particles are suspended in 
a liquid, they transfer momentum 
from and to each other via the liquid, hence their Brownian motion becomes 
correlated. This coupling of stochastic motion via hydrodynamic interactions 
is important in biological systems as well as microfluidic devices.

Recently Zimmermann and co-workers presented a series of articles on 
the correlations that are induced between two trapped colloidal particles 
by shear flow. They addressed the problem 
experimentally \cite{ZiehlZimmermann2009} and 
theoretically \cite{BammertZimmermann2010, HolzerZimmermann2010}.
The experimental system consisted of uncharged polystyrene beads suspended 
in water in a microfluidic device that allowed to create a linear shear 
profile. The particles were trapped by optical tweezers and their 
positions were recorded by with high speed camera. We will refer to the 
results of the experiments in section \ref{sec:TwoHS}.
In the theoretical work a Langevin model was solved. We briefly introduce 
the model and its solutions in section \ref{sec:EOM}.

Here we present a computer simulation study on the same problem. In section 
\ref{sec:SimDetails} we describe the simulation method, and in 
section \ref{sec:Results} we compare our results to the experimental and 
theoretical results.

\section{Equations of motion\label{sec:EOM}}
We briefly review the work of Zimmermann and co-workers
\cite{BammertZimmermann2010, HolzerZimmermann2010}:
The systems condsidered here are that of one and of two hard spherical particles trapped in harmonic potential wells and suspended in a shear flow with a linear velocity profile. Here and in the following the Brownian spheres have the same mass $M_i=M$ and the same effective radii $\varrho_i=\varrho$ with $i=1,2$. Their position vectors are denoted by ${\bi r}_i=\left(x_i,y_i,z_i\right)$. The isotropic potential wells $U_i\left({\bi r}_i\right)=\frac{k}{2}\left({\bi r}_i-{\bi q}_i\right)^2$ are located at ${\bi q}_1=\left(0,0,0\right)$ (single bead) and ${\bi q}_i=\left(\pm D/2,0,0\right)$ (pair of beads). Here $k$ is the strength and $D$ is the separation of the potentials. The potentials give rise to forces given by ${\bi F}^{\rm p}_i=-\nabla U_i$. As the colloids are surrounded by a fluid, they experience a force ${\bi F}^{\rm f}$ due to the friction with the fluid molecules. For small Reynolds numbers {\it Re} this force is proportional to the relative velocity of the colloid and the local fluid field, ${\bi F}^{\rm f}_i=-\zeta\left({\bi u}-\dot{\bi r}_i\right)$, where $\zeta=6\pi\eta\varrho$ is the Stokes friction coefficient and $\eta$ is the fluid viscosity. The linear velocity profile of the sheared suspension is given by ${\bi u}({\bi r})=\dot{\gamma}z\hat{\bi e}_x$ with the shear rate $\dot{\gamma}$. We are interested in the case of overdamped motion, i.e. the inertial part $M\ddot{\bi r}_i$ is negligible, and one can write down the Langevin equation
\begin{equation}\label{eq:Langevin}
	\dot{\bi r}_i = {\bi u}({\bi r}_i) + H_{ij}{\bi F}^{\rm p}_j + {\bi F}^{\rm s}_i
\end{equation}
with the mobility matrices
\begin{eqnarray}\label{eq:MobMatrix}
\eqalign{
	&H_{11} = H_{22} = \frac{1}{\zeta}I_{3\times3}, \\
 	&H_{12} = H_{21} = \frac{1}{\zeta}\frac{3\varrho}{4r_{12}}\left(I_{3\times3}+\frac{{\bi r}_{12}{\bi r}_{12}^{\rm T}}{r_{12}^2}\right).
}
\end{eqnarray}
Here, ${\bi r}_{12}:={\bi r}_1-{\bi r}_2$  is the distance vector between the two particles and its absolut value $r_{12}=\left|{\bi r}_{12}\right|$. The last contribution in eq.~\eref{eq:Langevin} is the stochastic force acting on particle $i$. For a fluid, in which orthogonal components of velocity fluctuations of the fluid molecules are uncorrelated, the following characteristics of ${\bi F}^{\rm s}_i$ can be assumed:
\begin{eqnarray}\label{eq:StochForces}
\eqalign{
	&\langle{\bi F}^{\rm s}_i(t) \rangle = 0, \\
	&\langle{\bi F}^{\rm s}_i(t){\bi F}^{\rm s}_j(t^\prime)\rangle = 2k_BTH_{ij}\delta\left(t-t^\prime\right).
}
\end{eqnarray}
One can rewrite eq.~\eref{eq:Langevin} in a more compact form (cf. \cite{BammertZimmermann2010}) by defining the vectors ${\bi R}:=\left({\bi r}_1,{\bi r}_2\right)$ and ${\bi Q}:=\left({\bi q}_1,{\bi q}_2\right)$ as well as the $6\times6$ matrices
\begin{equation}
	H := \Biggl(\begin{array}{cc}H_{11} & H_{21} \\ H_{12} & H_{22}\end{array}\Biggr),
\end{equation}
and ${\bi U}({\bi R}):=U{\bi R}$, where $U_{13}=U_{46}=\dot{\gamma}$ and all other $U_{kl}=0$,
so that it now reads as
\begin{equation}\label{eq:CompactLangevin}
	\dot{{\bi R}} = U{\bi R} + kH\left({\bi Q}-{\bi R}\right) + {\bi F},
\end{equation}
in which ${\bi F}$ satisfies \eref{eq:StochForces} with ${\bi F}_i^{\rm s}\to{\bi F}=\left({\bi F}_1^{\rm s},{\bi F}_2^{\rm s}\right)$ and $H_{ij}\to H$.
The coupled equations of motion \eref{eq:CompactLangevin} have been solved by Zimmermann and co-workers~\cite{BammertZimmermann2010,HolzerZimmermann2010}. Their solutions were given in terms of the particle fluctuations around the potential minima $\tilde{{\bi R}}(t)={\bi R}(t)-{\bi Q}$ as
\begin{equation}
	\tilde{{\bi R}}(t) = e^{-tM}\tilde{{\bi R}}(0) + \int_{0}^{t}{\rm d}t^\prime e^{(t^\prime -t)M}{\bi F},
\end{equation}
where $M:=kH-U$. They form the basis of the analytic calculation of the correlation functions $\langle\tilde{{\bi R}}(0)\tilde{{\bi R}}(t)\rangle$, where the brackets $\langle\ldots\rangle$ denote an ensemble average.

\section{Simulation details\label{sec:SimDetails}}
In order to simulate colloidal particles embedded in a sheared fluid environment, we used a combination of a simple molecular dynamics (MD) simulation and multiparticle-collision dynamics (MPCD), a mesoscopic solvent model to account for hydrodynamic interactions \cite{MalevanetsKapral1999}.

\subsection{Multiparticle-collision dynamics\label{sec:MPCD}}
The MPCD method we employ has been developed to solve the equations of hydrodynamics in a fluctuating solvent. $N$ point particles of the same mass $m$ are used for the transport of momentum through the system, while satisfying the conservation laws of mass, energy and momentum locally. The algorithm consists of two steps, namely free streaming interrupted by multiparticle collisions. In the streaming step, all fluid particles are propagated ballistically with their velocities ${\bi v}_k$, i.e. the particle positions are updated according to a time increment $h$ as
\begin{equation}\label{eq:Streaming}
	{\bi x}_k(t+h)={\bi x}_k(t)+h{\bi v}_k(t).
\end{equation}
After the time step $h$, the $N$ fluid particles are sorted into a lattice of cubic cells of size $a\times a\times a$, so that on average $\bar{n}$ particles are within each collision cell. Then, the particle velocities are rotated around the center of mass velocity $\bar{\bi v}$ in this cell,
\begin{equation}\label{eq:Collision}
	{\bi v}_k(t+h)=\bar{{\bi v}}(t)+\Omega(\alpha)\left[{\bi v}_k(t)-\bar{{\bi v}}(t)\right],
\end{equation}
where $\Omega(\alpha)$ is a rotation matrix corresponding to a fixed angle $\alpha$, which is generated randomly for each cell. Before a collision step is carried out, the collision cell grid is shifted by a randomly chosen vector with components taken from the interval $\left[-a/2,a/2\right]$, to ensure Galilean invariance \cite{IhleKroll2001}. As the transport coefficients like the dynamic viscosity $\eta$ can be expressed as functions of the simulation parameters in an analytical form \cite{IhleKroll2003a,IhleKroll2003b,PooleyYeomans2005}, one can tune their values to fullfill the conditions of overdamping  and low {\it Re} given in sec. \ref{sec:EOM}.\\
The coupling between the colloidal particles and the fluid particles is accomplished in the simplest manner, which means that a point particle with instantaneous velocity ${\bi V}$ and mass $M$ takes part in the collision step eq.~\eref{eq:Collision} within its cell with $\bar{{\bi v}}=\left(M{\bi V}+m\sum^n{\bi v}\right)/\left(M+nm\right)$.\\
In order to shear the fluid in our system, we confined the simulation box of size $L\times L\times L$ using two flat walls in $z$ direction at $\pm L/2$. Those were moved in $x$ direction with velocities $v^{(\pm)}_{\rm w}=\pm\dot{\gamma}L/2$. To reduce fluid slip at the walls we used a bounce-back rule, i.e. if a fluid particle hits a wall during the streaming step, its velocity is inverted in the rest frame of the wall (${\bi v}^\prime\to-{\bi v}^\prime$), where ${\bi v}^\prime:={\bi v}-{\bi v}^{(\pm)}_{\rm w}$. In combination, we used an algorithm prosposed in \cite{LamuraKroll2001}. Here, $N_{\rm pp}$ pseudo-particles are inserted into the cells, which are cut by the flat walls as a consequence of the random shift, so that in the partially filled cell the average number density is restored, $n+N_{\rm pp}=\bar{n}$.

\subsection{Molecular dynamics\label{sec:MD}}
In the MD part of our simulations, the positions of the colloidal particles were computed according to the Velocity Verlet algorithm with time step $\Delta t$ several times between two consecutive MPCD steps. The number of position updates is given as $h/\Delta t$. In the case of a pair of beads, we did not account for collisions between those. Hence we would expect deviations from the behaviour derived in sec.~\ref{sec:EOM} for small distances between the potential wells. As the ``colloidal'' beads also took part in the collision step as point particles, there is no physical volume associated with them, and therefore no geometrical radius can be defined.\\
\newline
For our simulations we chose the following parameters: the temperature was set to $k_BT=1$ and was kept constant on average using the thermostat proposed in ref.~\cite{HuangWinkler2010}. The collision cell size $a$ and the solvent particle mass $m$ were set to unity as well. The simulation box had dimensions $30\,a\times30\,a\times30\,a$. The mass of the colloids was set to $M=2m$. The rotation angle had the value $\alpha=130^\circ$. The two time steps of our hybrid dynamics simulations were $\Delta t=0.0001\tau$ and $h=0.01\tau$, both in units of $\tau:=\sqrt{ma^2/k_BT}$. The collision cells were occupied on average by $\bar{n}=10$ point particles. With those parameters, the dynamic viscosity was calculated as $\eta\approx82.2\,\sqrt{k_BTm}/a^2$. The value for the strength of the potential wells was $k=10\,k_BT/a^2$ and the rate with which the fluid is sheared was set to $\dot{\gamma}=0.04\,\tau^{\,-1}$. With those values, a Reynolds number of $\mathcal{O}({\mathit Re})\approx0.1\ll2300$ results. Furthermore, the ratio $\zeta/m$, which is a measure for the damping in the system, is of the order of $\mathcal{O}\left({\zeta/m}\right)\approx10^2\gg1$. Hence, the trapped particles perform an overdamped motion embedded in a laminar fluid flow.

\section{Results\label{sec:Results}}
In the following we will present our results for both the auto-correlations (AC) and the cross-correlations (CC) in the random displacements of Brownian particles embedded in a liquid. The simulations were done in a quiescent as well as in a sheared fluid. We focus on fluctuations parallel ($x$) and perpendicular ($z$) to the direction of the shear flow. The correlation data was measured as the average of $10^5-10^6$ time series. Error bars are given by the standard deviation of the data sets.

\subsection{Single sphere\label{sec:OneHS}}
\begin{figure}
\centering
\includegraphics[scale=0.35]{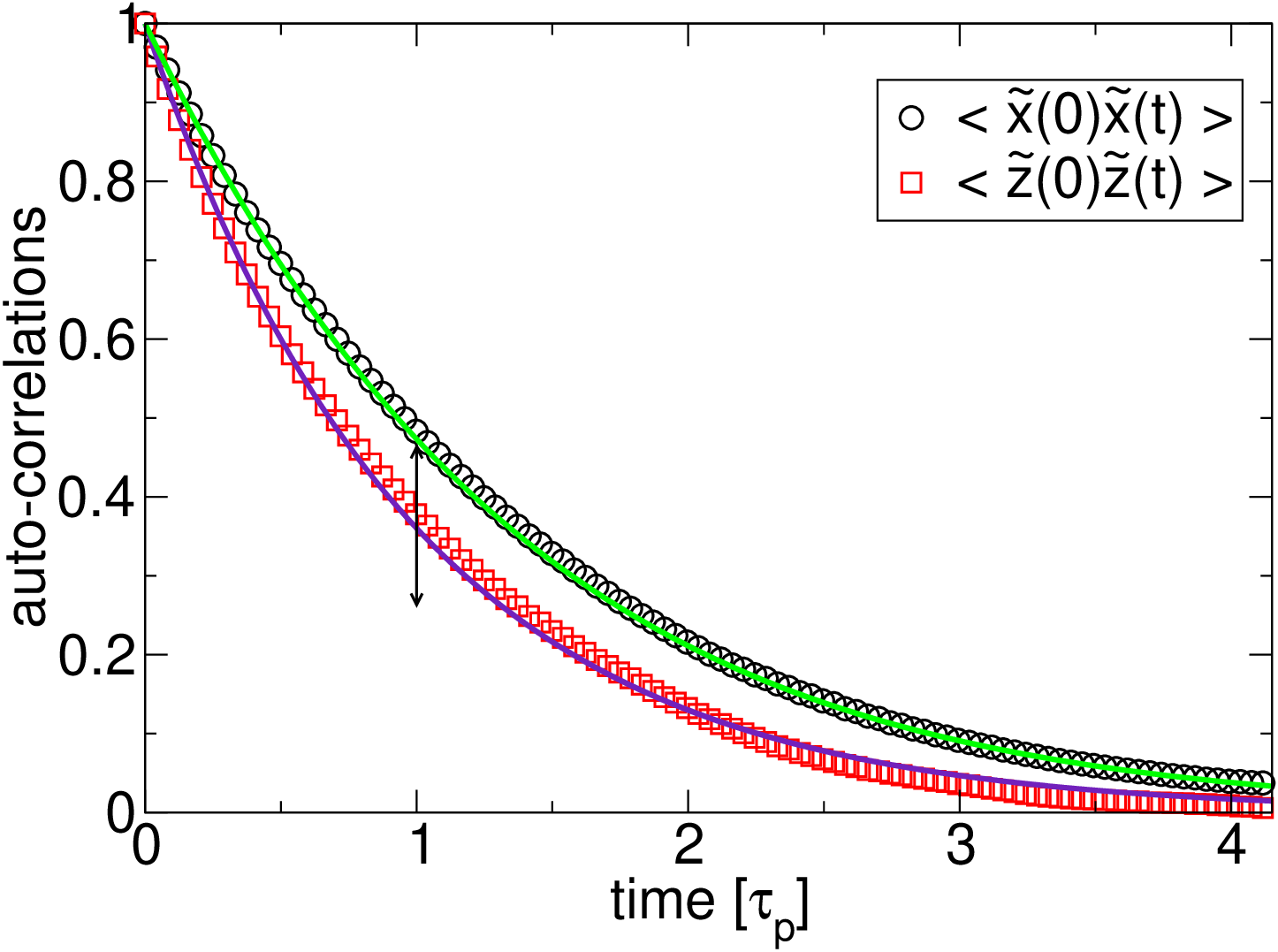}
\caption{Auto-correlation of particle position for a single particle in the direction of shear flow (simulation: black circles, theory: green line) and perpendicular to it (simulation: red squares, theory: blue line). The arrow denotes the size of the error bars.}
\label{fig:AutoCorr_OneHS_WS}
\end{figure}

\begin{figure}
\centering
\includegraphics[scale=0.35]{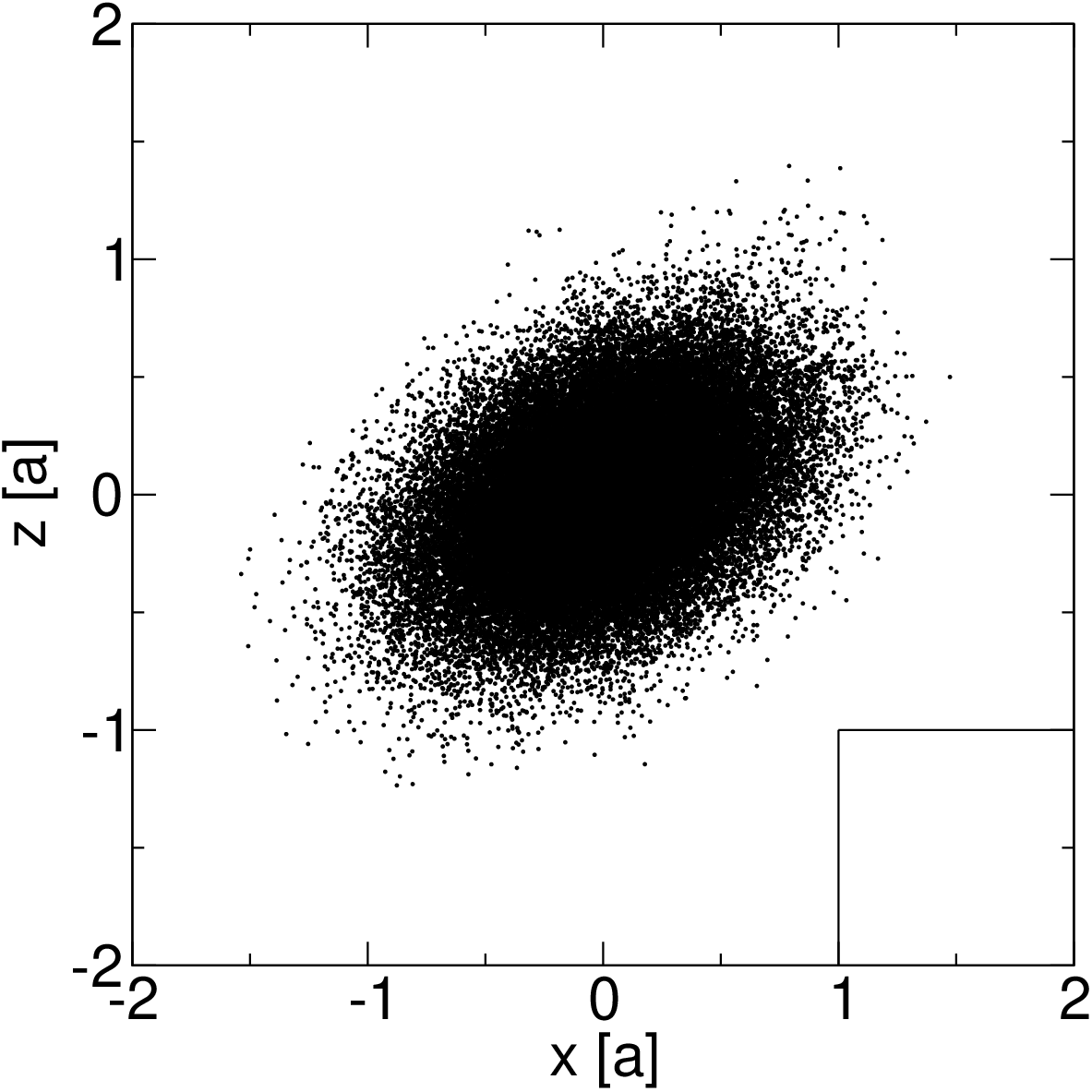}
\caption{Particle position distribution in the shear plane for a single particle in sheared flow. The box denotes the size of a collision cell $a\times a$.}
\label{fig:Pos_Distribution}
\end{figure}

\begin{figure}
\centering
\includegraphics[scale=0.35]{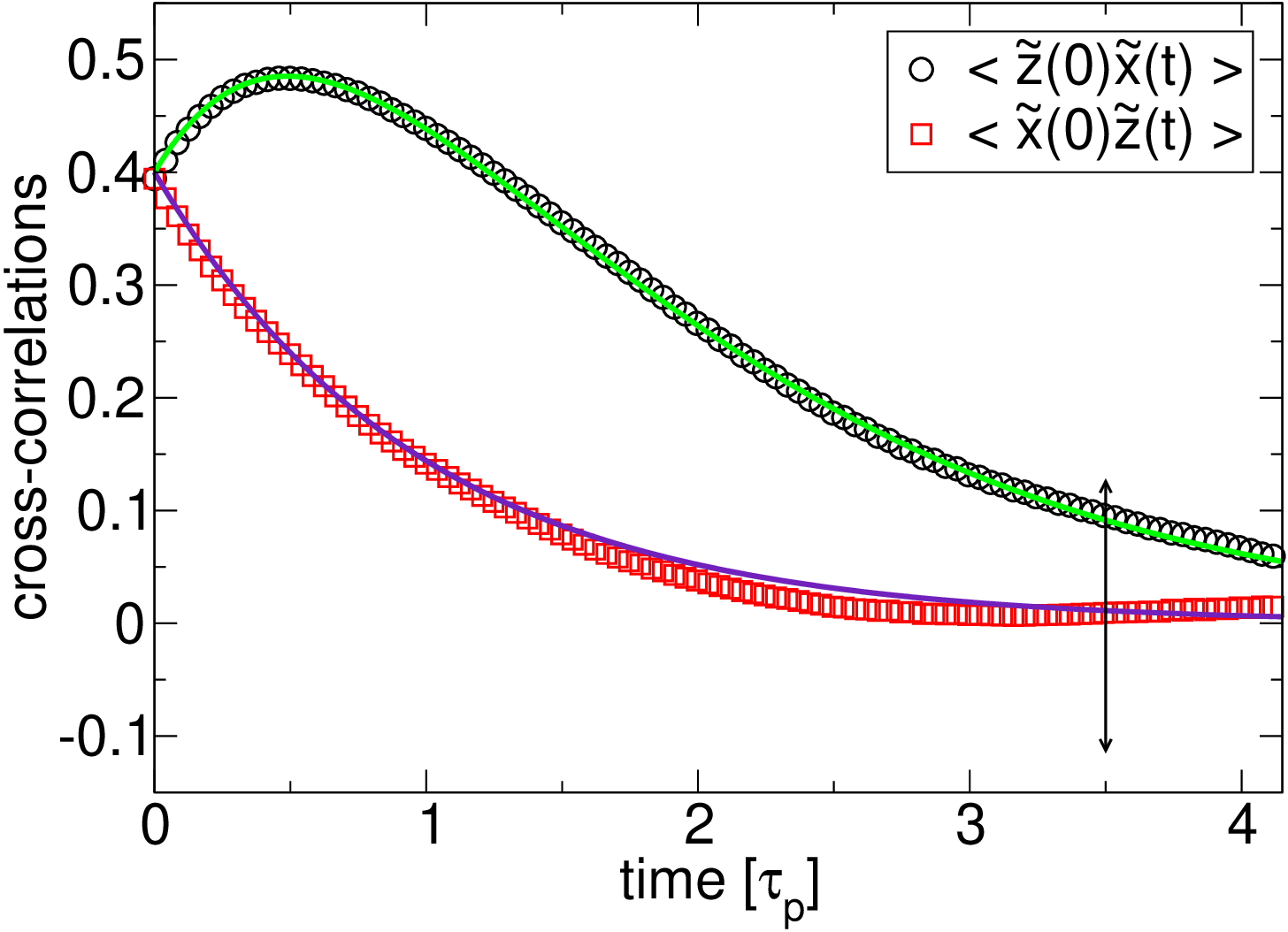}
\caption{Cross-correlations of particle position for a single particle in shear flow. The arrow denotes the size of the error bars.}
\label{fig:CrossCorr_OneHS_WS}
\end{figure}

Given the defining properties of the stochastic forces (eq.~\eref{eq:StochForces}), the Langevin equation eq.~\eref{eq:Langevin} for a single particle embedded in a fluid without flow leads to an isotropic AC function, which decays exponentially with a relaxation time $\tau_{\rm p}=\zeta/k$
\begin{equation}\label{eq:OneB_NS_AC}
	\langle \tilde{x}(0)\tilde{x}(t)\rangle = \langle \tilde{z}(0)\tilde{z}(t)\rangle = \frac{k_BT}{k}e^{-t/\tau_{\rm p}}.
\end{equation}
Using this equation, we determined the relaxation time from our simulation 
data as $\tau_{\rm p}=23.7\,\tau$.
As the fluctuations of orthogonal fluid velocity components do not couple for $\dot{\gamma}=0$, the CCs vanish, $\langle\tilde{x}(0)\tilde{z}(t)\rangle = \langle\tilde{x}(t)\tilde{z}(0)\rangle = 0$. \\
The situation changes if the colloidal bead is exposed to a sheared fluid. Then the AC in the shear direction is modified to
\begin{equation}\label{eq:OneB_WS_AC}
	\langle \tilde{x}(0)\tilde{x}(t)\rangle = \frac{k_BT}{k}\left[1+\frac{{\rm Wi}^2}{2}\left(1+\frac{t}{\tau_{\rm p}}\right)\right]e^{-t/\tau_{\rm p}},
\end{equation}
while $\langle\tilde{z}(0)\tilde{z}(t)\rangle$ equals the one of eq.~\eref{eq:OneB_NS_AC}. Here, eq.~\eref{eq:OneB_WS_AC} is given in terms of the Weissenberg number ${\rm Wi}:=\dot{\gamma}\tau_{\rm p}=0.95$. Both correlation functions are shown in figure \ref{fig:AutoCorr_OneHS_WS}.\\
\newline
Now, if there is a finite shear rate, the CCs are not zero anymore and fluctuations of orthogonal components couple under the influence of the linear velocity profile of the surrounding fluid. This can be visualized by the distribution of the particle positions in the shear plane, as it is depicted in figure \ref{fig:Pos_Distribution}. Here the distribution, which would be spherical in the case of a quiescent fluid, assumes an elliptical shape. As it was shown in ref.~\cite{HolzerZimmermann2010}, the characteristic dimensions of the ellipse, as the ratio between the principle axes $w/l$ and the inclination angle $\phi$, are connected to ${\rm Wi}$ via
\begin{eqnarray}\label{eq:EllipseDims}
\eqalign{
	\tan\phi &= \frac{1}{2}\left[\sqrt{4+{\rm Wi}^2}-{\rm Wi}\right], \\
	w/l &= \left(\frac{\sqrt{4+{\rm Wi}^2}-{\rm Wi}}{\sqrt{4+{\rm Wi}^2}+{\rm Wi}}\right)^{1/2}.
}
\end{eqnarray}
From the data in figure \ref{fig:Pos_Distribution}, we found $\phi=32.3\,^\circ$ and $w/l=0.622$. Using equations \eref{eq:EllipseDims}, those lead to the Weissenberg numbers ${\rm Wi}=0.95$ and ${\rm Wi}=0.98$, respectively, both in good agreement with the value calculated above. Also, in the simulation the CCs do not vanish any more. They show a behaviour which satisfies the equations
\begin{eqnarray}\label{eq:OneB_WS_CC}
\eqalign{
	\langle \tilde{x}(t)\tilde{z}(0)\rangle &= \frac{k_BT}{k}\frac{{\rm Wi}}{2}\left(1+2\frac{t}{\tau_{\rm p}}\right)e^{-t/\tau_{\rm p}}, \\
	\langle \tilde{x}(0)\tilde{z}(t)\rangle &= \frac{k_BT}{k}\frac{{\rm Wi}}{2}e^{-t/\tau_{\rm p}}.
}
\end{eqnarray}
Both functions are plotted in figure \ref{fig:CrossCorr_OneHS_WS}. There, the initial increase of $\langle \tilde{x}(t)\tilde{z}(0)\rangle$ reflects the fact, that finite fluctuations $\tilde{z}(0)\neq0$ are carried away by the shear flow in $x$-direction before the initial displacement starts to relax. The 
analytical predictions are in excellent agreement with the simulation data (no fit parameters).

\subsection{Two hard spheres\label{sec:TwoHS}}
\begin{figure}
\centering
\includegraphics[scale=0.35]{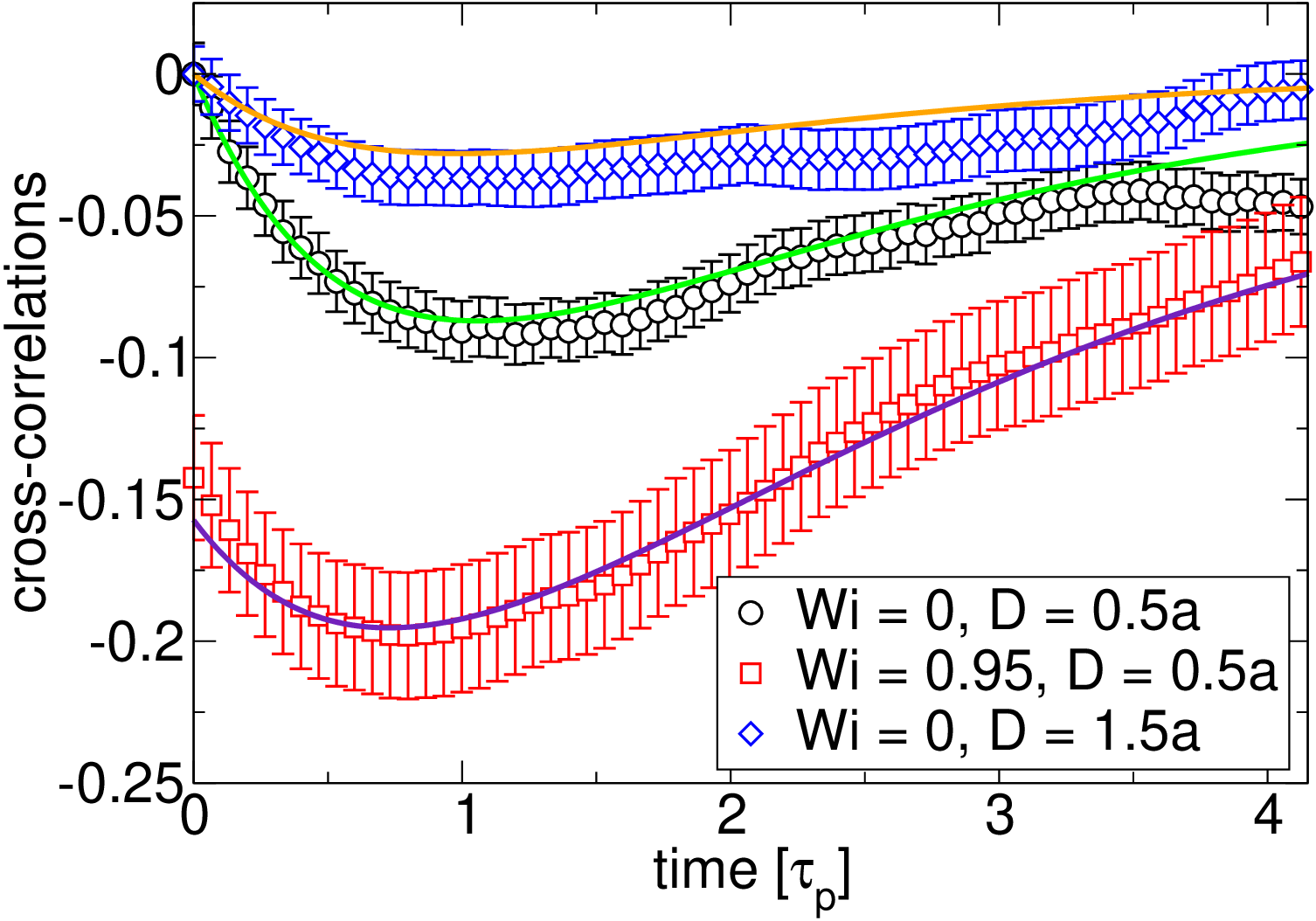}
\caption{Cross-correlations of fluctuations parallel to the shear direction, $\langle\tilde{x}_1(0)\tilde{x}_2(t)\rangle$, between two beads for unsheared (simulation: black circles, blue diamonds, theory: green and orange line) and sheared system (simulation: red squares, theory: blue line).}
\label{fig:CrossCorr_TwoHS_NS_WS}
\end{figure}

\begin{figure}
\centering
\includegraphics[scale=0.35]{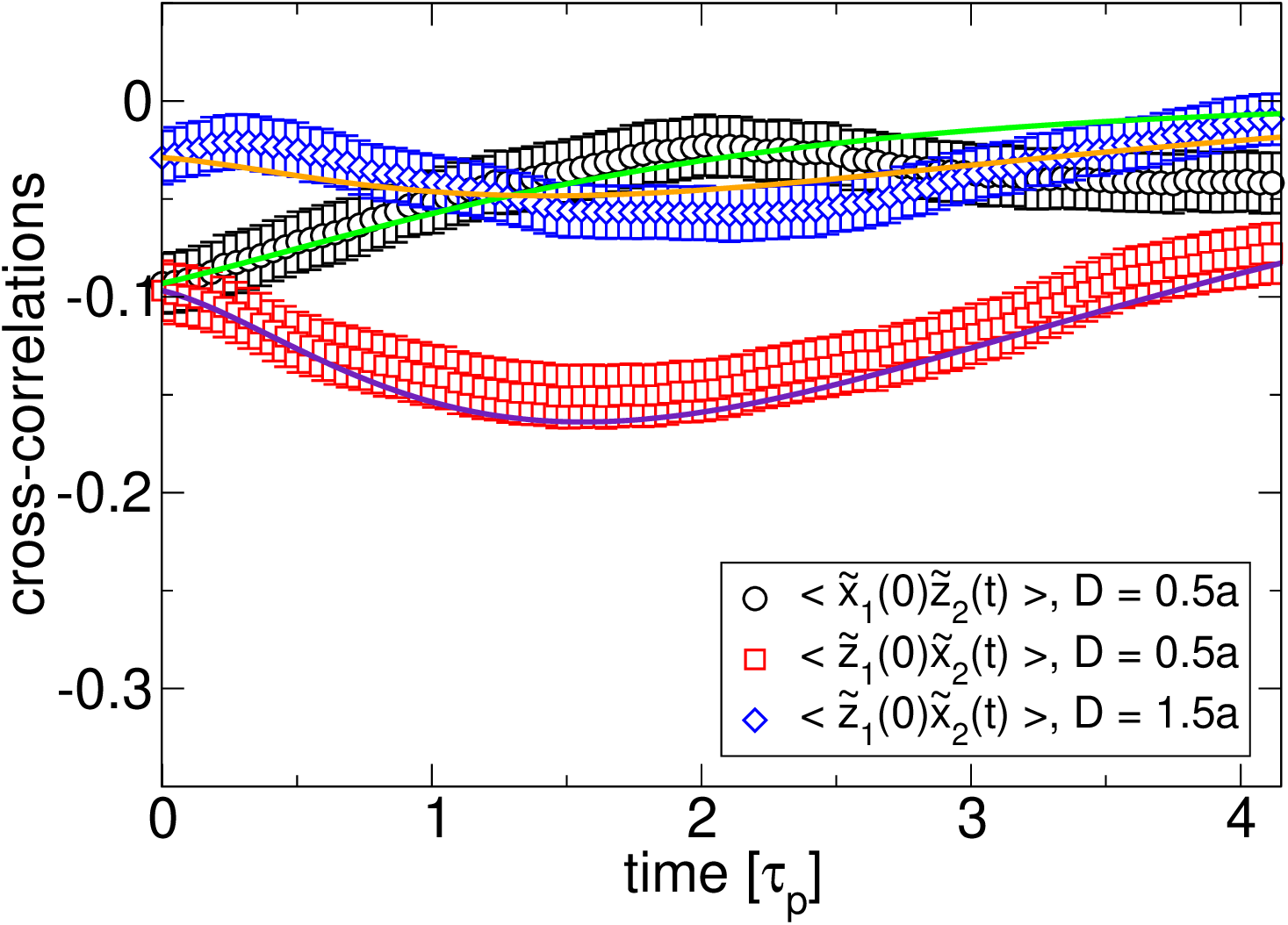}
\caption{Cross-correlations of fluctuations in perpendicular directions between two beads for $D=0.5a$ (black circles, green line and red squares, indigo line) and $D=1.5a$ (blue diamonds and orange line).}
\label{fig:CrossCorr_TwoHS_WS}
\end{figure}

In the case of two trapped particles, not only correlations between fluctuations in different directions of each colloid are of interest, but also inter-particle correlations like $\langle \tilde{x}_1(t)\tilde{x}_2(0)\rangle$. Those are functions of the distance between the potential minima $D=\left|{\bi q}_{12}\right|$, where ${\bi q}_{12}:={\bi q}_1-{\bi q}_2$ is the connection vector of the two potential wells. In order to compare our results with analytical expressions, one has to be in a regime where the hydrodynamic radii $\varrho$ of the particles are small compared to $D$. To obtain a value for $\varrho$, we used the definitions of the relaxation time $\tau_{\rm p}=\zeta/k$ and the friction coefficient, from which it follows that $\varrho=\tau_{\rm p}k/6\pi\eta$. In our simulations the hydrodynamic radius roughly equals $0.15\,a$.\\
In this section we will follow the nomenclature of ref.~\cite{BammertZimmermann2010} and introduce the four relaxation rates:
\begin{eqnarray}\label{eq:DefRelaxationTimes}
\eqalign{
	\lambda_1 &= \frac{1+2\mu}{\tau_{\rm p}},\qquad\lambda_3 = \frac{1-2\mu}{\tau_{\rm p}}, \\
	\lambda_2 &= \frac{1+\mu}{\tau_{\rm p}},\qquad\,\,\,\lambda_4 = \frac{1-\mu}{\tau_{\rm p}},
}
\end{eqnarray}
with the parameter $\mu:=3\varrho/4D$. To illustrate the meaning of the $\lambda_i$, let us consider a situation in which the two particles, after experiencing independent stochastic kicks, are pulled back into the minima of their potential wells, respectively. This relaxation process can be decomposed into two modes for each spacial direction, namely a parallel and an anti-parallel translation of the particles. As the hydrodynamic interaction between the spheres depends on their relative motion, restoring is accelerated in the parallel and damped in the anti-parallel case. Here, the modes for the two directions perpendicular to ${\bi q}_{12}$ coincide, so that only the above relaxation rates are left over.\\
A previous experimental study \cite{MeinersQuake1999} with two trapped Brownian particles in a quiescent fluid showed anti-correlations between random fluctuations along same directions. These anti-correlations are a function of $\mu$. In addition, the functions are modified under the influence of a linear shear profile in a similar manner as the AC function for fluctuations in shear direction in the one-particle case \eref{eq:OneB_WS_AC}. Including this shear flow correction, which is proportional to ${\rm Wi}^2$, $\langle\tilde{x}_1(0)\tilde{x}_2(t)\rangle$ is given by
\begin{eqnarray}\label{eq:TwoB_NS_CC}
\eqalign{
	\langle\tilde{x}_1(0)\tilde{x}_2(t)\rangle = &\,\frac{1}{4}\left(e^{-\lambda_1 t}-e^{-\lambda_3 t}\right)\\
					&\,+\frac{{\rm Wi}^2}{4\mu}\left(\frac{-\left(1+\mu\right)e^{-\lambda_1 t}}{6\mu^2+7\mu+2}+\frac{e^{-\lambda_2 t}}{2+3\mu}\right)\\ 
					&\,+\frac{{\rm Wi}^2}{4\mu}\left(\frac{-\left(1+\mu\right)e^{-\lambda_3 t}}{6\mu^2-7\mu+2}+\frac{e^{-\lambda_4 t}}{2-3\mu}\right)
}
\end{eqnarray}
The simulation data for the CC functions and eq.~\eref{eq:TwoB_NS_CC} are plotted in figure \ref{fig:CrossCorr_TwoHS_NS_WS} for an unsheared system (${\rm Wi} = 0$) with two different $D$ and a sheared (non-zero ${\rm Wi}$) system. Again, the agreement with the analytical predictions is very good. We note that not only the depths of the minima differ but also the position of the strongest anti-correlation changes when the particles are embedded in a sheared environment.\\
In the data for ${\rm Wi}=0$ and $D=0.5a$ (black circles in figure \ref{fig:CrossCorr_TwoHS_NS_WS}) there is an additional decrease of the CC function at long times, which is not covered by the error bars. A similar behaviour can be found in the data of ref.~\cite{MeinersQuake1999} for small distances between the laser potentials. The origin and the physical relevance of this effect need further investigation.\\
Finally, we will briefly discuss inter-particle correlations between perpendicular fluctuation directions, namely $\langle\tilde{x}_1(0)\tilde{z}_2(t)\rangle$ and $\langle\tilde{z}_1(0)\tilde{x}_2(t)\rangle$. The behaviour of the latter can be understood as a combination of the inter-particle anti-correlation of fluctuations in shear direction and the fact, that a fluctuation in $z$-direction of a single particle is followed by a positive motion in $x$-direction (cf. equation \eref{eq:OneB_WS_CC})
\begin{equation*}
‪	\tilde{x}_2\;\overset{\rm{antiCC}}{\longrightarrow}\;\tilde{x}_1=\tilde{x}_1\left(\tilde{z}_1\right)\;\overset{\rm{shear\;flow}}{\longrightarrow}\;\tilde{z}_1,
\end{equation*}
which produces an anti-correlation in $\langle\tilde{z}_1(0)\tilde{x}_2(t)\rangle$. The analytical expressions for the two inter-particle CC functions are given by
\begin{eqnarray}\label{eq:TwoB_WS_CC}
\eqalign{
	\langle\tilde{x}_1(0)\tilde{z}_2(t)\rangle = &\,\frac{\rm Wi}{4}\left(\frac{e^{-\lambda_2 t}}{2+3\mu}-\frac{e^{-\lambda_4 t}}{2-3\mu}\right),\\
	\langle\tilde{z}_1(0)\tilde{x}_2(t)\rangle = &\,\frac{\rm Wi}{4}\left(e^{-\lambda_2 t}-e^{-\lambda_4 t}\right)\\
					&\,-\frac{\rm Wi}{2\mu}\left(\frac{\left(1+\mu\right)e^{-\lambda_1 t}}{2+3\mu}+\frac{\left(1-\mu\right)e^{-\lambda_3 t}}{2-3\mu}\right).
}
\end{eqnarray}
These expressions are compared to the simulation data in 
figure \ref{fig:CrossCorr_TwoHS_WS}. Again, they agree within the 
errorbars.

\subsection{Comparison with experiment\label{sec:CompSimExp}}
\begin{figure}
\centering
\includegraphics[scale=0.35]{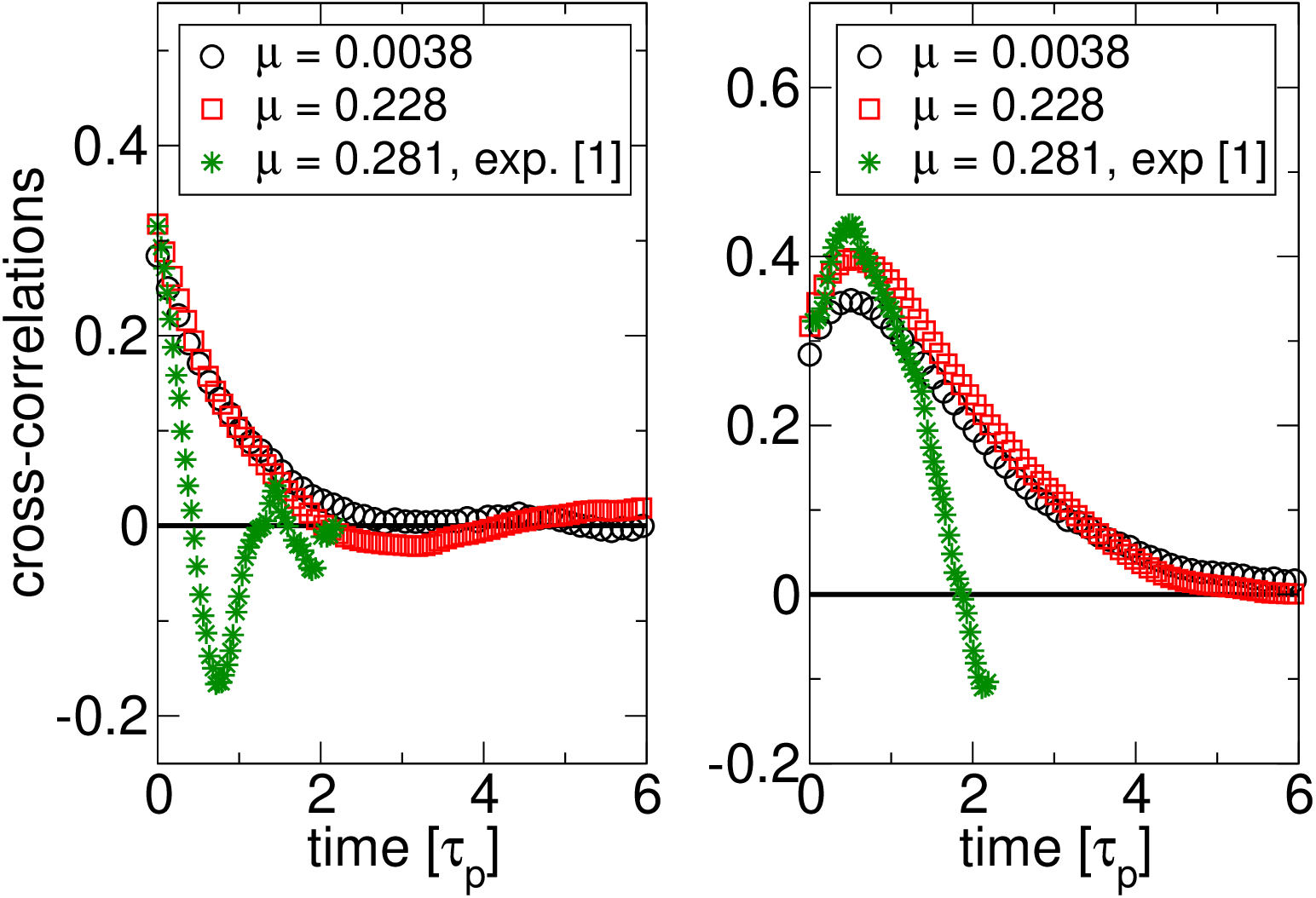}
\caption{Single-particle cross correlation functions $\langle \tilde{x}(0)\tilde{z}(t)\rangle$ (left panel) and $\langle \tilde{x}(t)\tilde{z}(0)\rangle$ (right panel). Comparison of simulation data for $\mu=0.0038$ (black circles) and $\mu=0.228$ (red squares) with experimental results (green crosses).}
\label{fig:CrossCorr_CompExp_OneB}
\end{figure}

In this section we compare our results to the experimental data from 
ref.~\cite{HolzerZimmermann2010}. The green stars in 
fig.~\ref{fig:CrossCorr_CompExp_OneB} are the experimental values for the single particle cross correlation functions $\langle \tilde{x}(0)\tilde{z}(t)\rangle$ (left panel) and $\langle \tilde{z}(0)\tilde{x}(t)\rangle$ (right panel). There are clear differences with respect to the simulation results in fig.~\ref{fig:CrossCorr_OneHS_WS}. $\langle \tilde{x}(0)\tilde{z}(t)\rangle$ shows a pronounced minimum, while the simulated curve drops monotonically (red squares in fig.~\ref{fig:CrossCorr_OneHS_WS}), and the decay of the experimental 
$\langle \tilde{z}(0)\tilde{x}(t)\rangle$ is much faster than of the simulation data (black circles in fig.~\ref{fig:CrossCorr_OneHS_WS}).
 
In contrast to our simulation (in which there was only one particle), in the experiment the single-particle CCs 
were obtained using a system that contained two particles. Hence, the single-particle correlations of perpendicular fluctuations are influenced by presence of the second particle. For small potential well distances this can lead to deviations from the analytical description given in \cite{BammertZimmermann2010}, which is only valid for small $\mu$, i.e. large distances $D$.
In the PhD thesis of A.~Ziehl \cite{ZiehlThesis} it is suggested that these deviations are possibly the origin of the additional minimum in $\langle \tilde{x}(0)\tilde{z}(t)\rangle$. To test this hypothesis, we also analysed the CCs for reasonably large $\mu$ with a second particle present. The simulation data is plotted in figure \ref{fig:CrossCorr_CompExp_OneB}. Here, the functions were renormalized to account for the different Weissenberg number of ${\rm Wi}_{\rm exp}\approx0.62$. Our data shows the development of a minimum with increasing $\mu$ which supports the hypothesis. However, the minimum is much less pronounced than the one in the experimental data. Hence there might be other sources for this effect.\\

\begin{figure}
\centering
\includegraphics[scale=0.35]{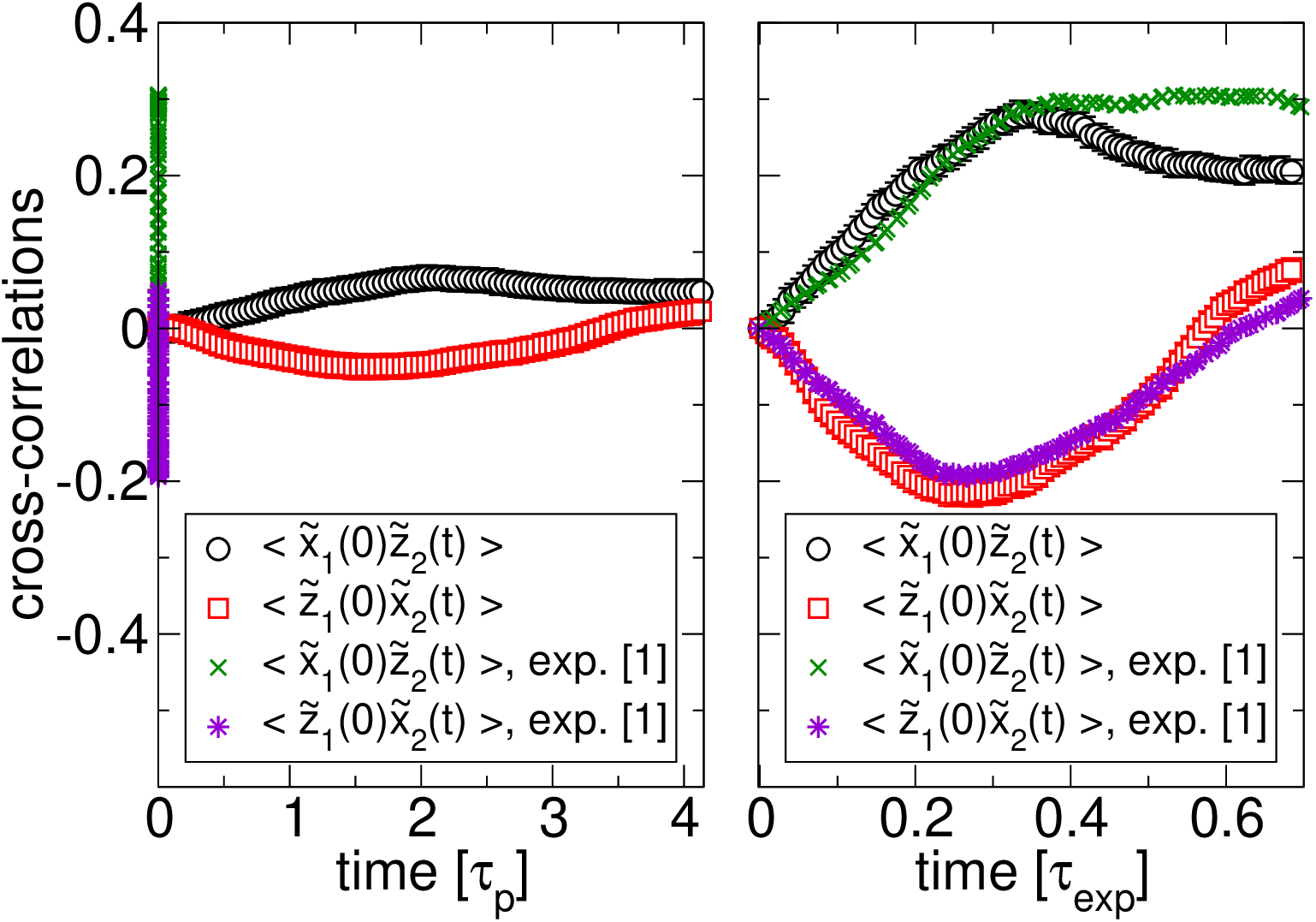}
\caption{Inter-particle cross correlation functions. Comparison of simulation data (black circles and red squares)  with experimental results (green crosses and violet stars). Left panel shows data with time rescaled by $\tau_{\rm p}$. Right panel shows results with time rescaled by $\tau_{\rm exp}$, given by minimum of experimental $\langle\tilde{z}_1(0)\tilde{x}_2(t)\rangle$, and correlation strength rescaled with respect to different Weissenberg numbers.}
\label{fig:CrossCorr_CompExp_TwoB}
\end{figure}

Fig.~\ref{fig:CrossCorr_CompExp_TwoB} shows the experimental data for the two particle correlation functions. There is a clear difference in time-scales between simulation and experiment (left panel).
This is suspected to be due to deviations in the prodcution process of the microfluidic device, which cause changes in the trapping potential \cite{Schneider2012}. To compare our results of the inter-particle CC functions, we renormalized the time scale to match the experimental relaxation time $\tau_{\rm exp}$, which is given by the position of the minimum of $\langle\tilde{z}_1(0)\tilde{x}_2(t)\rangle$ and which differs from the relaxation time used before in ref.~\cite{HolzerZimmermann2010}.  We again took into account the different ${\rm Wi}$. After these transformations have been applied, the data sets show very good agreement (right panel).


\section{Summary}
We have presented a computer simulation study of the hydrodynamic 
interactions between two colloids in a sheared fluid. In particular, we have
tested the hypothesis that the recently observed non-monotonic behaviour in the cross-correlation function of the position of a single particle 
is due to the presence of a second particle. We found that a 
second particle produces this effect, but it might not be the full explanation.
Another source could be the rotational motion of the colloids, which we neglected in the simulation.
   
\section*{Acknowledgements}
Funding by the German Research foundation (DFG) within the Sfb TR6 project 
D5 and by the Fonds National de la Recherche Luxembourg within the AFR 
scheme (PHD-09-177)
is greatfully acknowledged.

\end{document}